\documentclass{article}
\usepackage{ijcai21}

\usepackage[font=small,labelfont=bf]{caption}
\usepackage[belowskip=-9pt,aboveskip=3pt]{caption} 

\usepackage{times}
\usepackage{soul}
\usepackage{url}
\usepackage[utf8]{inputenc}
\usepackage{graphicx}
\usepackage{amsmath}
\usepackage{amsthm}
\usepackage{booktabs}
\usepackage{algorithm}
\usepackage{algorithmic}
\usepackage{color}
\usepackage[hang,flushmargin]{footmisc} 
\usepackage{pifont}
\usepackage{sidecap}

\newcommand{\tabincell}[2]{\begin{tabular}{@{}#1@{}}#2\end{tabular}}

\usepackage{xspace}
\makeatletter
\DeclareRobustCommand\onedot{\futurelet\@let@token\@onedot}
\def\@onedot{\ifx\@let@token.\else.\null\fi\xspace}
\def\eg{\emph{e.g}\onedot} 
\def\ie{\emph{i.e}\onedot} 
 
\def\etc{\emph{etc}\onedot}

\makeatother

\usepackage{multirow}
\usepackage{colortbl}
\definecolor{tabgray}{rgb}{0.85,0.85,0.85}
\definecolor{top1}{rgb}{1.0, 0.6, 0.6} 
\definecolor{top2}{rgb}{0.98, 0.91, 0.71}
\definecolor{top3}{rgb}{0.91, 1.0, 1.0}
\definecolor{top1-2}{rgb}{1.0, 0.66, 0.66} 
\definecolor{top1-3}{rgb}{1.0, 0.72, 0.72} 
\definecolor{top1-4}{rgb}{1.0, 0.78, 0.78} 
\definecolor{top1-5}{rgb}{1.0, 0.84, 0.84} 
\definecolor{top1-6}{rgb}{1.0, 0.90, 0.90} 
\definecolor{top1-7}{rgb}{1.0, 0.96, 0.96} 

\definecolor{citecolor}{RGB}{65,105,225}
\usepackage[pagebackref=true,breaklinks=true,colorlinks,citecolor=citecolor,bookmarks=false]{hyperref}

\title{AVA: Adversarial Vignetting Attack against Visual Recognition}

\author{
Binyu Tian$^1$\and
Felix Juefei-Xu$^2$\and 
Qing Guo$^3$\footnote{Qing Guo and Xiaohong Li are the corresponding authors ( \href{mailto:qing.guo@ntu.edu.sg}{qing.guo@ntu.edu.sg} and \href{mailto:xiaohongli@tju.edu.cn}{xiaohongli@tju.edu.cn}).}\and
Xiaofei Xie$^3$\and
Xiaohong Li$^1$\footnotemark[1]\and
Yang Liu$^{3}$\\ 
\affiliations
$^1$ College of Intelligence and Computing, Tianjin University, China\\
$^2$ Alibaba Group, USA\\
$^3$ Nanyang Technological University, Singapore\\
}

\begin{document}

\maketitle

\begin{abstract}
Vignetting is an inherit imaging phenomenon within almost all optical systems, showing as a radial intensity darkening toward the corners of an image.
Since it is a common effect for the photography and usually appears as a slight intensity variation, people usually regard it as a part of a photo and would not even want to post-process it. 
Due to this natural advantage, in this work, we study the vignetting from a new viewpoint, \ie, \textit{adversarial vignetting attack (AVA)}, which aims to embed intentionally misleading information into the vignetting and produce a natural adversarial example without noise patterns.
This example can fool the state-of-the-art deep convolutional neural networks (CNNs) but is imperceptible to human.
To this end, we first propose the \textit{radial-isotropic adversarial vignetting attack (RI-AVA)} based on the physical model of vignetting, where the physical parameters (\eg, illumination factor and focal length) are tuned through the guidance of target CNN models. 
To achieve higher transferability across different CNNs, we further propose \textit{radial-anisotropic adversarial vignetting attack (RA-AVA)} by allowing the effective regions of vignetting to be radial-anisotropic and shape-free. 
Moreover, we propose the geometry-aware level-set optimization method to solve the adversarial vignetting regions and physical parameters jointly.
We validate the proposed methods on three popular datasets, \ie, DEV, CIFAR10, and Tiny ImageNet, by attacking four CNNs, \eg, ResNet50, EfficientNet-B0, DenseNet121, and MobileNet-V2, demonstrating the advantages of our methods over baseline methods on both transferability and image quality.
\end{abstract}

\section{Introduction}


\begin{figure}[t]
	\centering
	\includegraphics[width=\columnwidth]{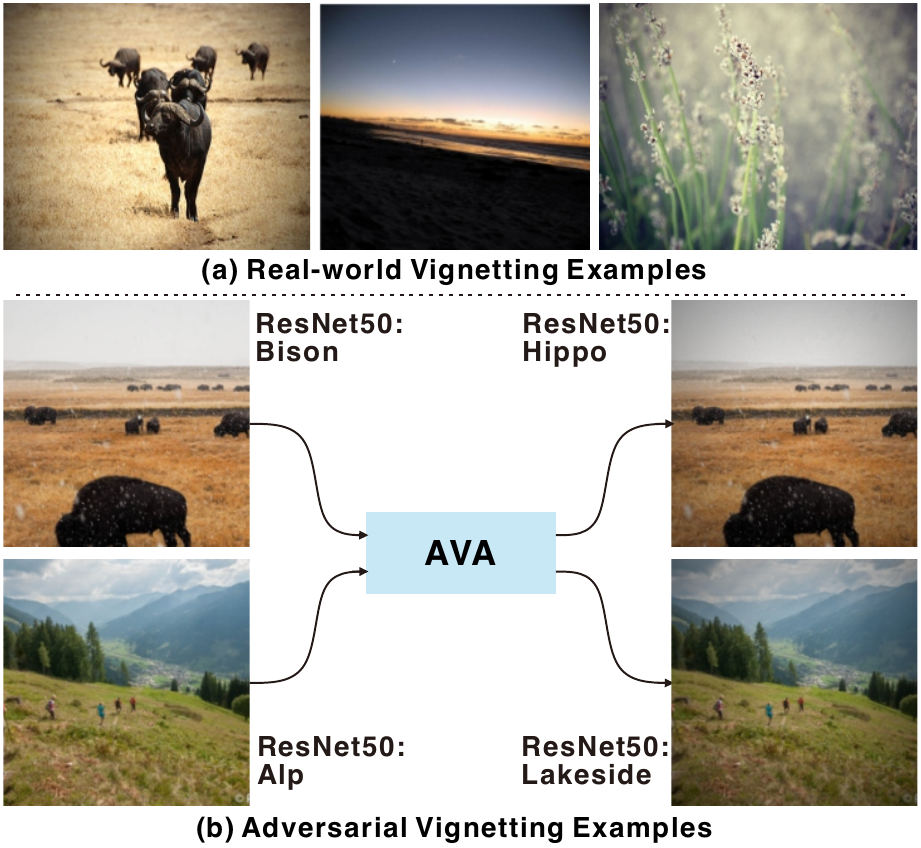}
	\caption{(a) shows three real vignetting images captured by cameras. (b) shows the adversarial examples produced by our adversarial vignetting attack (AVA), fooling the SOTA CNN ResNet50 with imperceptible property due to the realistic vignetting effects.}
	\label{fig:teaser}
\end{figure}

In photography, image vignetting is a common effect as a result of camera settings or lens limitations. It shows up as a gradually darkened transparent ring-shape mask towards the image border with continuous reduction of the image brightness or saturation \cite{gonzalez2004digital}. 
Vignetting often naturally occurs during the photo taking process. As categorized by \cite{ray2002applied}, there are the following three main causes of vignetting for digital imaging: (1) mechanical vignetting, (2) optical vignetting, and (3) natural vignetting.
Both \emph{mechanical vignetting} and \emph{optical vignetting} are somehow caused by the blockage of light. For example, the mechanical vignetting is caused by light emanated from off-axis scene being partially blocked by external objects such as lens hoods, and the optical vignetting is usually caused by the multiple element lens setting where the effective lens opening for off-axis incident light can be reduced. 
On the other hand, \emph{naturally vignetting} is not due to light blockage, but rather by the law of illumination falloff where the light falloff is proportional to the 4-th power of the cosine of the angle at which the light particles hit the digital sensor. 
Sometimes, vignetting can also be applied on the digital image as an artistic post-processing step to draw people's attention to the center portion of the photograph as depicted in Fig.~\ref{fig:teaser}(a). 


Therefore, image vignetting can be capitalized to ideally hide adversarial attack information in a stealthy way for its ubiquity and naturalness in digital imaging. In this work, we propose a novel and stealthy attack method called the adversarial vignetting attack (AVA) that aims at embedding intentionally misleading information into the vignetting and producing a natural adversarial example without noise patterns, as depicted in Fig.~\ref{fig:teaser}(b). By first mathematically and physically model the image vignetting effect, we have proposed the radial-isotropic adversarial vignetting attack (RI-AVA) and the physical parameters such as the illumination factors and the focal length are tuned through the guidance of the target CNN models under attack. Next, by further allowing the effective regions of vignetting to be radial-anisotropic and shape-free, our proposed radial-anisotropic adversarial vignetting attack (RA-AVA) can achieve much higher transferability across different CNN models. 
Moreover, we have proposed the level-set-based optimization method, that is geometry-aware, to solve the adversarial vignetting regions and physical parameters jointly.



Through extensive experiments, we have validated the effectiveness of the proposed methods on three popular datasets, \ie, DEV \cite{devDB}, CIFAR10 \cite{krizhevsky2009learning}, and Tiny ImageNet \cite{tinyImageNet}, by attacking four CNNs, \eg, ResNet50 \cite{he2016deep}, EfficientNet-B0 \cite{tan2019efficientnet}, DenseNet121 \cite{huang2017densely}, and MobileNet-V2 \cite{sandler2018mobilenetv2}. We have successfully demonstrated the advantages of our methods over strong baseline methods especially on transerferbility and image quality. To the best of our knowledge, this is the very first attempt to formulate stealthy adversarial attack by means of image vignetting and showcase both the feasibility and the effectiveness through extensive experiments.

\section{Related Work}
{\bf Adversarial noise attacks.}
Adversarial noise attacks aim to fool DNNs by adding imperceptible perturbations to the images. One of the most popular attack methods, \ie, fast gradient sign method (FGSM) \cite{goodfellow2014explaining}, involves only one back propagation step in the process of calculating the gradient of cost function, enabling simple and fast adversarial example generation. \cite{kurakin2016adversarial} proposes an improved version of FGSM, known as basic iteration method (BIM), which heuristically search for examples that are most likely to fool the classifier. \cite{dong2018boosting} proposes a broad class of momentum-based iterative algorithms to boost adversarial attacks. By integrating the momentum term into the iterative process for attacks, it can stabilize update directions and escape from poor local maxima during the iterations. \cite{dong2019evading} further proposes a translation-invariant attack method to generate more transferable adversarial examples against the defense models. 
Adversarial noise attack will generate patterns that do not exist in reality, and our method is an early method of applying patterns that may be generated in natural optical systems to attacks.

{\bf Other adversarial attacks.}
In addition to traditional adversarial noise attacks, there are some methods that focus on sparse or real-life patterns.
\cite{croce2019sparse} proposes a new attack method to generate adversarial examples aiming at minimizing the $l_0$-distance to the original image. It allows pixels to change only in region of high variation and avoiding changes along axis-aligned edges, resulting in almost non-perceivable adversarial examples.
\cite{wong2019wasserstein} proposes a new threat model for adversarial attacks based on the Wasserstein distance. The resulting algorithm can successfully attack image classification models, bringing traditional CIFAR10 models down to 3\% accuracy within a Wasserstein ball with radius 0.1.
\cite{bhattad2019unrestricted} introduces ``unrestricted'' perturbations that manipulate semantically meaningful image-based visual descriptors (\eg, color and texture) to generate effective and photorealistic adversarial examples.
\cite{guo2020watch} proposes an adversarial attack method that can generate visually natural motion-blurred adversarial examples. 
Along similar lines, non-additive noise based adversarial attacks that focus on producing realistic degradation-like adversarial patterns have emerged in several recent studies, such as adversarial rain \cite{zhai2020s} and haze \cite{gao2021advhaze}, adversarial exposure for medical analysis \cite{cheng2020adversarial} and co-saliency detection problems \cite{gao2020making}, adversarial bias field in medial imaging \cite{tian2020bias}, as well as adversarial denoising \cite{cheng2020pasadena} and face morphing \cite{wang2020amora}.
These methods apply patterns that may be produced in reality such as color and texture to attack, but ignore the patterns that are generated naturally in the optical systems, which are also vitally important.

{\bf Vignetting correction methods.}
Research into vignetting correction has a long history.
The Kang-Weiss model \cite{kang2000can} is established to simulate the vignetting effect. It proves that it is possible to calibrate a camera using just a flat, textureless Lambertian surface and constant illumination.
\cite{zheng2008single} proposes a method for robustly determining the vignetting function in order to remove the vignetting given only a single image.
\cite{goldman2010vignette} further proposes a method to remove the vignetting from the images without resolving ambiguities or the previously known scale and gamma ambiguities. 
These works on correcting the vignetting effect provides a basis for us to model the vignetting effect. Inspired by these work, we will capitalize the vignetting effect as a means of an adversarial attack.


\section{Adversarial Vignetting Attack (AVA)}
Vignetting effect is related to numerous factors, \eg, angle-variant light across camera sensor, intrinsic lens characteristics, and physical occlusions. 
There are several works studying how to model the vignetting including empirical-based \cite{goldman2010vignette,yu2004practical} and physical-based methods \cite{asada1996photometric,kang2000can}. 
In particular, Kang and Weiss \cite{kang2000can} proposes a physical-based method that models vignetting via physically meaningful parameters (\eg, off-axis illumination, light path obstruction, tilt effects), allowing better understanding of the influence from real-world environments (\eg, camera setups or physical occlusion) to the final results.
In this section, we start from the physical model of vignetting effects \cite{kang2000can} and propose two adversarial vignetting attacks based on this model with an level-set-based optimization.

\begin{SCfigure*}
	\centering
	\includegraphics[width=0.8\textwidth]{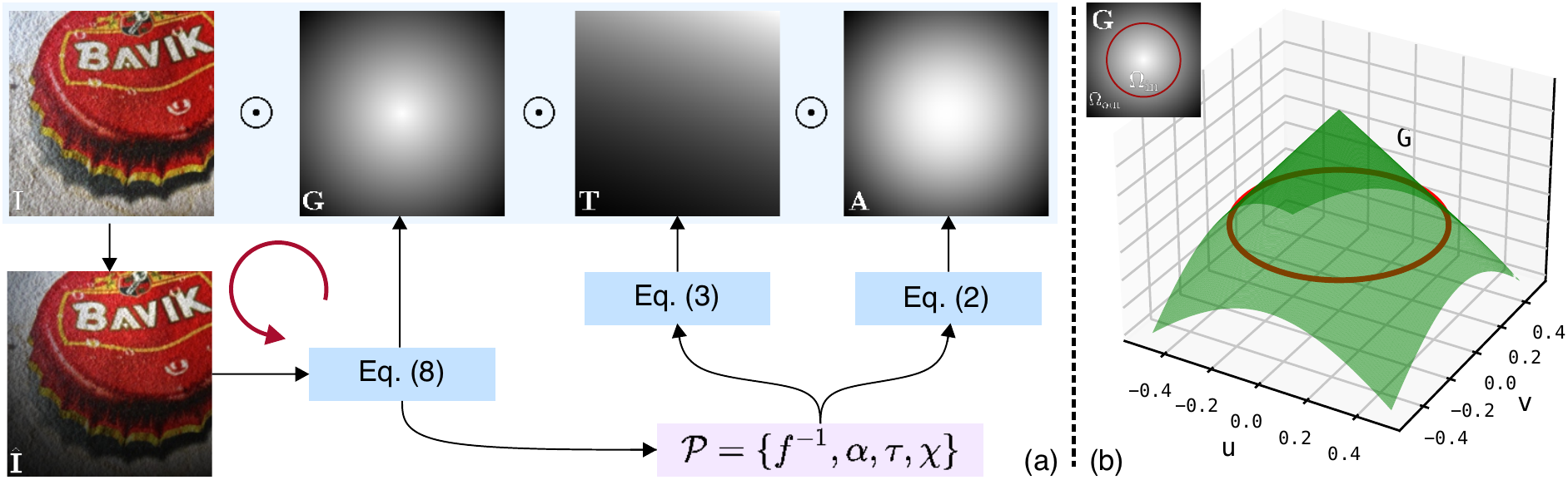}
	\caption{(a) shows the whole process of RA-AVA. (b) shows the 3D surface of the initialized $\mathbf{G}$. The red line is the curve splitting the image to 2 parts, \ie, $\Omega_\text{in}$ and $\Omega_\text{out}$.}
	\label{fig:frameworks}
\end{SCfigure*}

\subsection{Physical Model of Vignetting}
Given a clean image $\mathbf{I}$, we aim to simulate the vignetting image\footnote{\scriptsize{Throughout the paper, the term `vignetting image' refers to a photographic image that exhibits the vignetting effect to some degree.}} via $\hat{\mathbf{I}} = \mathbf{I}\odot\mathbf{V}$ where $\mathbf{V}$ is a matrix having the same size with $\mathbf{I}$ and represents the vignetting effects, and $\odot$ denotes the pixel-wise multiplication.
We model the vignetting from three aspects, \ie, off-axis illumination factor $\mathbf{A}$, geometric factor $\mathbf{G}$, and a tilt factor $\mathbf{T}$ \cite{kang2000can}. 
All three factors are pixel-wise and have the same size with the original image.
Then, the vignetting effects can be also represented as
%
\begin{align}\label{eq:physical}
\mathbf{V} = \mathbf{A}\odot\mathbf{G}\odot\mathbf{T}.
\end{align}
%

Intuitively, $\mathbf{A}$ describes the phenomenon of the illumination in the image that is darkened with distance away from the image center \cite{kang2000can}, defined as
%
\begin{align}\label{eq:vigA}
\mathbf{A} = \frac{1}{ \big(1+(\frac{\mathbf{R}}{f})^2 \big)^2},
\end{align}
%
where $f$ is the effective focal length of the camera, and $\mathbf{R}$ is a fixed matrix and denotes the distance of each pixel to the principal point, \ie, the image center with the coordinate as $(u,v)=(0,0)$ if the lens distortion does not exist. 

The matrix $\mathbf{G}$ represents the vignetting caused by the off-axis angle projection from the scene to the image plane \cite{tsai1987versatile}, and is approximated by
%
\begin{align}\label{eq:vigG}
\mathbf{G} = 1-\alpha\mathbf{R},
\end{align}
%
where $\alpha$ is a scalar deciding the geometry vignetting degree. 

The matrix $\mathbf{T}$ defines the effects of camera tilting to image plane and the $i$-th element is formulated as
%
\begin{align}\label{eq:vigT}
\mathbf{T}[i] = \cos{\tau} \Big(1+\frac{\tan{\tau}}{f}(u_i\sin{\chi}-v_i\cos{\chi})^2 \Big),
\end{align}
%
where $\chi$ and $\tau$ are tilt-related parameters determining the camera pose w.r.t. a scene/object. Please find more details in \cite{kang2000can}. 

With this physical model, we aim to study the effects of vignetting from the viewpoint of adversarial attack, \eg, how to actively tune the vignetting-related parameters, \ie, $f$, $\alpha$, $\tau$, and $\chi$, to let the simulated vignetting images to fool the state-of-the-art CNNs easily? To this end, we represent the vignetting process as a simple function, \ie,
%
\begin{align}\label{eq:vigfunc}
\hat{\mathbf{I}} = \text{vig}(\mathbf{I},\mathcal{P})=\mathbf{I}\odot\mathbf{V},
\end{align}
%
where $\mathcal{P}=\{f^{-1},\alpha,\tau,\chi\}$. Then, we propose the radial-isotropic adversarial vignetting attack (RI-AVA).

\subsection{Radial-Isotropic AVA}
Given a clean image $\mathbf{I}$ and a pre-trained CNN $\phi$, we aim to tune the $\mathcal{P}=[f^{-1},\alpha,\tau,\chi]$ under a norm ball constraint for each parameter. 
\begin{align}\label{eq:riava_adv_obj}
\arg\max_{\mathcal{P}}~&J(\phi(\text{vig}(\mathbf{I},\mathcal{P}),y) + \lambda_{f}|f|_2 - \lambda_{\alpha}|\alpha|_2, \nonumber\\
& \mathrm{~~subject~to~~} \forall \rho\in\mathcal{P}, |\rho|_\text{p}\le \epsilon_{\rho},
\end{align}
where the first term $J(\cdot)$ is the image classification loss under the supervision of the annotation label $y$, the second and third terms encourage the focal length to be larger and geometry coefficient $\alpha$ to be smaller. As a result, the clean image $\mathbf{I}$ would not be changed significantly. 
Besides, $\epsilon_{\rho}$ denotes the ball bound under $L_\text{p}$ for the parameter $\rho$. Here, we use the infinite norm. 
We can optimize the objective function by gradient descent-based methods, that is, we calculate the gradient of the loss function with respect to all parameters in $\mathcal{P}$ and update them to realize the gradient-based attacks like existing adversarial noise attacks \cite{BIM_2016_ICLRW,guo2020watch}.

Since this method equally tunes the pixels on the same radius to the image center, we name it as \textit{radial-isotropic adversarial vignetting attack (RI-AVA)}. 
Nevertheless, by tuning only four scalar physical-related parameters to realize attack, this method can study the robustness of CNN to realistic vignetting effects but it is hard to realize intentional attacks with high attack success rate and high transferability across different CNNs. To fill this gap, we further propose the radial-anisotropic adversarial vignetting attack (RA-AVA) by extending the geometry vignetting $\mathbf{G}$, allowing the each element of $\mathbf{G}$ to be independently tuned.

\subsection{Radial-Anisotropic AVA}
To enable more flexible vignetting effects, we allow $\mathbf{G}$ to be tuned independently in an element-wise way and redefine the objective function in Eq.~\eqref{eq:riava_adv_obj} to jointly optimize $\mathbf{G}$ and $\mathcal{P}$.
Specifically, for the matrix $\mathbf{G}$, we split it into two parts with a closed curve $\mathcal{C}$ centered at the principal point. 
On the one hand, we desire the region of $\mathbf{G}$ inside $\mathcal{C}$ (\ie, $\Omega_\text{in}$) to be similar with the physical function defined by Eq.~\eqref{eq:vigG}, making the simulated image look naturally. 
%
%
In contrast, we also want all elements of $\mathbf{G}$ to be flexibly tuned according to the adversarial classification loss, leading to high attack success rate.
In particular, the vignetting effects let pixels in the outside region darker than the ones in the $\Omega_\text{in}$.
Hence, embedding adversarial information into this region is less risky to be perceived. 
Overall, we define a new objective function to tune $\mathbf{G}$, $\mathcal{C}$, and $\mathcal{P}$ jointly
%
\begin{align}\label{eq:raava_adv_obj}
&\arg\max_{\mathbf{G},\mathcal{P},\mathcal{C}}~J(\phi(\text{vig}(\mathbf{I},\mathcal{P}),y) - \lambda_{g}\sum_{i\in \Omega_\text{in}}|(\mathbf{G}[i]-\mathbf{G}_0[i])|_2  \nonumber\\
& ~~~~ + \lambda_{f}|f|_2 - \lambda_{\alpha}|\alpha|_2, \mathrm{~~subject~to~~} \forall \rho\in\mathcal{P}, |\rho|_\text{p}\le \epsilon_{\rho},
\end{align}
%
where $\mathbf{G}_0=1-\alpha\mathbf{R}$, and the region $\Omega_\text{in}$ is determined by the curve $\mathcal{C}$. Note that, we tune $\mathcal{C}$ along its inward normal direction to let shape and area of $\Omega_\text{out}$ be changed according to the adversarial classification loss (\ie, $J(\cdot)$). For example, when the  $\Omega_\text{out}$ becomes larger and then less pixels (\ie, $\Omega_\text{in}$) are constrained by the second term of Eq.~\eqref{eq:raava_adv_obj}, we have more flexibility to tune the pixels in the image to reach high attack success rate. 
We can solve Eq.~\eqref{eq:raava_adv_obj} by regarding it as a curve evolution problem \cite{kass1988snakes} since the curve $\mathcal{C}$ is an optimization variable. Nevertheless, this method can hardly handle topological changes of the moving front, such as splitting and merging of $\mathcal{C}$ \cite{KIMIA1992438}. Inspired by the works \cite{guo2018frequency} for curve optimization, we propose to regard the $\mathbf{G}$ as the level-set function of the curve and solve Eq.~\eqref{eq:raava_adv_obj} via our geometry-aware level-set optimization.

\begin{figure}[t]
	\centering
    \includegraphics[width=1\columnwidth]{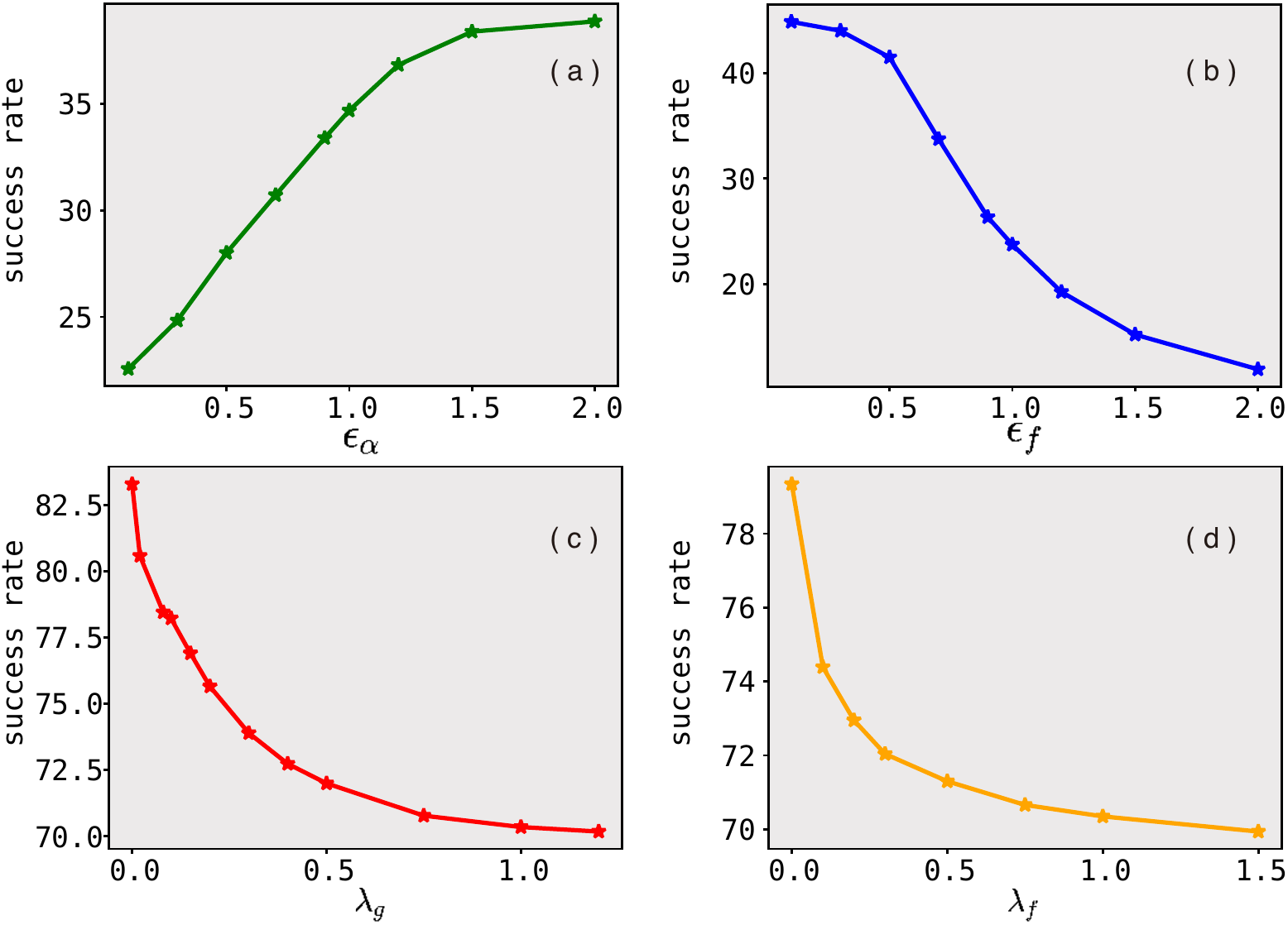}
	\caption{ Influences of hyper-parameters. (a) and (b) shows success rate w.r.t. different upper bounds of $f$ (\ie, $\epsilon_f$) and $\alpha$ (\ie, $\epsilon_\alpha$) under the RI-AVA method. We also show the success rate w.r.t. different  $\lambda_g$ and $\lambda_{f}$ in (c) and (d), respectively, under the RA-AVA method.}
	\label{fig:f_alf_trend}
\end{figure}

\subsection{Geometry-aware Level-Set Optimization}
To optimize Eq.~\eqref{eq:raava_adv_obj}, we first initialize the geometry vignetting, \ie, $\mathbf{G}=\mathbf{G}_0=1-\alpha\mathbf{R}$, which is a distance-related matrix and we reformulate it as a level-set function $\text{g}(u_i,v_i)=\mathbf{G}[i]=1-\alpha\sqrt{u_i^2+v_i^2}$. Intuitively, the function $\text{g}(\cdot)$ takes the coordinates of a position on the image plane as inputs and outputs a value that decreases as the coordinates become larger, leading to a 3D surface.
With such a function, we can define the curve $\mathcal{C}$ as the $z$-level-set of $\text{g}(\cdot)$, \ie, $\mathcal{C}=\{(u_i,v_i)|z=\text{g}(u_i,v_i)\}$, that is, $\mathcal{C}$ is the cross section of the 3D surface at level $z$.
Then, we can define the region inside the curve as $\Omega_\text{in}=\{(u_i,v_i)|z<\text{g}(u_i,v_i)\}$, which can be reformulated it as the function of $\text{g}$ by the Heaviside function (\ie, $\text{H}(\cdot)$) \cite{chan2001active}, that is, we have $\Omega_\text{in}=\{(u_i,v_i)|\text{H}(g(u_i,v_i))>z\}~\text{or}~\text{H}(\mathbf{G})$.
Finally, we can reformulate Eq.~\eqref{eq:raava_adv_obj} as
%
\begin{align}\label{eq:raava_adv_obj_levelset}
&\arg\max_{\mathbf{G},\mathcal{P}}~J(\phi(\text{vig}(\mathbf{I},\mathcal{P}),y) - \lambda_{g}\|(\mathbf{G}-\mathbf{G}_0)\odot\text{H}(\mathbf{G})\|^2_2  \nonumber\\
&~~~ +\lambda_{f}|f|_2 - \lambda_{\alpha}|\alpha|_2,  \mathrm{~~subject~to~~} \forall \rho\in\mathcal{P}, |\rho|_\text{p}\le \epsilon_{\rho}.
\end{align}
%
Since the Heaviside function is differentiable we can optimize the objective function via gradient descent. Compared with Eq.~\eqref{eq:raava_adv_obj}, the proposed new formulation only contains two terms that should be optimized, \ie, $\mathbf{G}$ and $\mathcal{P}$, making the optimization more easier.
In practice, we can calculate the gradient of $\mathbf{G}$ and $\mathcal{P}$ w.r.t. to the objective function and use the signed gradient descent to optimize $\mathbf{G}$ and $\mathcal{P}$ as is done in \cite{BIM_2016_ICLRW}.


\subsection{Implementation Details}

We show the whole process in Fig.~\ref{fig:frameworks}(a). Specifically, given a clean image $\mathbf{I}$ and a DNN $\phi(\cdot)$, we summarize the workflow of our attacking algorithm in the following steps:  \ding{182} Initialize the parameters $\mathcal{P}=\{f^{-1},\alpha,\tau,\chi\}=\{1,0,0,0\}$, the geometry vignetting matrix $\mathbf{G}$ as $1-\alpha\mathbf{R}$, and the distance matrix $\mathbf{R}$ via $\mathbf{R}[i]=\sqrt{u_i^2+v_i^2}$.
\ding{183} Calculate the illumination-related matrix $\mathbf{A}$ via Eq.~\eqref{eq:vigA}, and the camera tilting-related matrix $\mathbf{T}$ via Eq~\eqref{eq:vigT}.
\ding{184} At the $t$-th iteration, calculate the gradient of $\mathbf{G}_t$, $\mathcal{P}_t$ with respect to the objective function Eq.~\eqref{eq:raava_adv_obj_levelset} and obtain $\nabla_{\mathbf{G}_t}$ and $\{\nabla_{\rho_t}|\rho_t\in\mathcal{P}_t\}$. 
\ding{185} Update $\nabla_{\mathbf{G}_t}$ and $\mathcal{P}_t$ with their own step sizes. 
\ding{186} Update $t=t+1$ and go to the step three for further optimization until it reaches the maximum iteration or $\text{vig}(\mathbf{I},\mathcal{P})$ fools the DNN.
In the experimental parts, we set our hyper-parameters as follows: we set the stepsize of $f,\alpha,\tau,\chi$ and 
$\nabla_{\mathbf{G}_t}$ as 0.0125, 0.0125, 0.01, 0.01 and 0.0125, respectively. We set the number of iterations to be 40 and $z$ of the level-set method to be 1.0. We set ${p}$ to be $\infty$, and set the $\epsilon$ of $f^{-1}$, $\alpha$, $\tau$, and $\chi$ as 0.5, 0.5, $\pi/6$, and $\pi/6$. In addition, we set $\lambda_{f}$, $\lambda_{g}$ and $\lambda_{\alpha}$ all to be 1. In Section~\ref{sec:exp}, we will carry out experiments to evaluate the effect of different hyper-parameters. And we do not choose the hyper-parameters for the highest success rate when compared with baseline attacks, but rather set the parameters that can balance the high success rate and good image quality.

\section{Experimental Results}\label{sec:exp}

\begin{figure}[t]
	\centering
	\includegraphics[width=1\columnwidth]{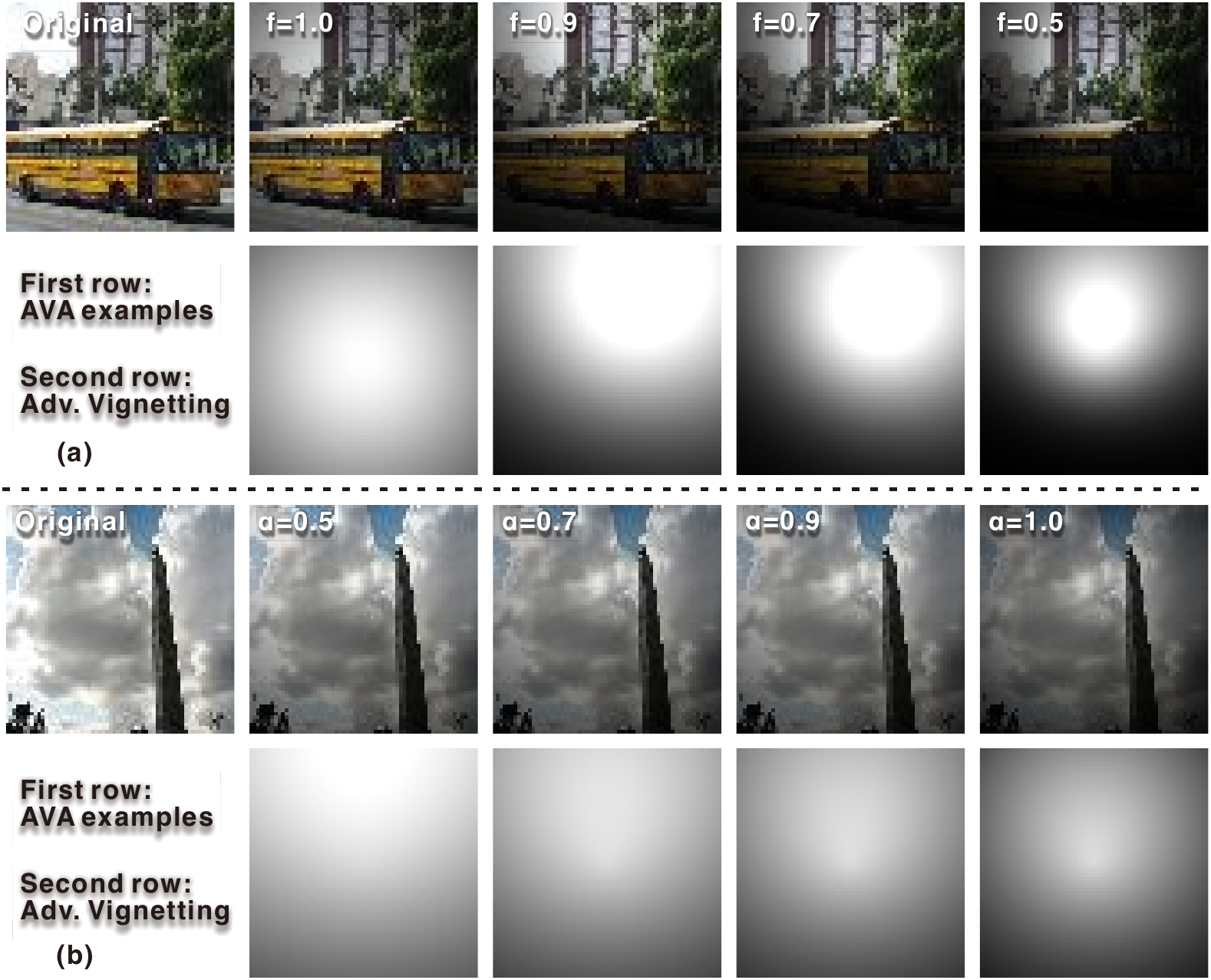}
	\caption{ Visualization results of different ball bound for $f$ and $\alpha$.
	}
	\label{fig:f-alf}
\end{figure}
\begin{table*}[t]
\centering
\setlength{\tabcolsep}{4pt}
{
    \resizebox{1.0\linewidth}{!}{
    {
	\begin{tabular}{l|l|ccc|ccc|ccc|ccc}
		
		\toprule
		
		\rowcolor{tabgray}\cellcolor {white}&\multicolumn{1}{c|}{Crafted from} & \multicolumn{3}{c|}{ResNet50} & \multicolumn{3}{c|}{EfficientNet}& \multicolumn{3}{c|}{DenseNet} & \multicolumn{3}{c}{MobileNet} \\
		\cmidrule(r){2-2} \cmidrule(r){3-5} \cmidrule(r){6-8} \cmidrule(r){9-11}\cmidrule(r){12-14}
		\rowcolor{tabgray}
		\cellcolor {white}\multirow{-2}{*}{} &\tabincell{c}{ Succ Rate\\ \& Metrics}
		 &  Succ Rate & BRISQUE & NIQE
		
		 &  Succ Rate & BRISQUE & NIQE
		
		 &  Succ Rate & BRISQUE & NIQE
		   
		 &  Succ Rate & BRISQUE & NIQE\\
		
		\midrule
		\multirow{8}{*}{\rotatebox[origin=c]{90}{DEV}} 
	& MIFGSM & 99.78  & 20.93  & 39.42  & 99.89  & 19.12  & 39.38  & 100.00  & 21.17  & 40.19  & 100.00  & 22.79  & 42.19  \\
      & CW    &  100.00  & 17.49  & 48.36  & 100.00  & 17.81  & 48.36  & 100.00  & 17.48  & 48.53  & 100.00  & 17.33  & 48.51  \\
      & TIMIFGSM & 96.23  & 18.34  & 45.86  & 98.56  & 18.59  & 46.15  & 98.34  & 18.48  & 45.60  & 98.94  & 18.55  & 45.99  \\
      & Wasserstein & 14.21  & 20.91  & 51.63  & 32.78  & 20.44  & 51.30  & 16.50  & 20.67  & 51.79  & 13.87  & 20.02  & 51.58  \\
      & cAdv  & 81.59  & 18.44  & 51.48  & 88.27  & 18.45  & 51.38  & 77.85  & 18.43  & 51.46  & 78.73  & 18.44  & 51.36  \\
      \cline{2-14}
      & \textbf{RI-AVA} & 9.36  & 19.78  & 48.33  & 14.95  & 20.06  & 48.24  & 13.07  & 20.10  & 48.17  & 20.68  & 20.32  & 48.12  \\
      & \textbf{RA-AVA}  & 96.77  & 21.33  & 46.92  & 98.34  & 22.81  & 47.02  & 99.22  & 20.89  & 46.54  & 99.18  & 21.20  & 46.66  \\

		\midrule
		\multirow{8}{*}{\rotatebox[origin=c]{90}{CIFAR10}} & MIFGSM & 80.78  & 41.86  & 42.04  & 96.67  & 41.90  & 42.53  & 79.03  & 41.40  & 41.97  & 97.87  & 41.56  & 42.14  \\
      & CW    & 100.00  & 41.43  & 41.36  & 100.00  & 41.66  & 41.06  & 100.00  & 41.34  & 41.42  & 100.00  & 41.46  & 41.01  \\
      & TIMIFGSM & 38.80  & 41.66  & 40.64  & 38.64  & 41.54  & 40.59  & 34.82  & 41.48  & 40.62  & 59.77  & 41.44  & 40.67  \\
      & Wasserstein & 80.27  & 45.45  & 44.47  & 74.73  & 44.81  & 44.09  & 80.62  & 43.26  & 42.05  & 66.43  & 43.45  & 42.87  \\
      & cAdv  & 12.78  & 41.42  & 40.78  & 21.24  & 41.55  & 40.78  & 11.88  & 41.32  & 40.84  & 17.28  & 41.40  & 40.84  \\
      
      \cline{2-14}
      & \textbf{RI-AVA} & 6.17  & 40.54  & 40.28  & 9.53  & 39.90  & 40.34  & 6.73  & 40.57  & 40.28  & 12.52  & 39.73  & 40.36  \\
		
      & \textbf{RA-AVA} & 35.95  & 33.56  & 38.05  & 74.15  & 28.29  & 35.39  & 45.80  & 31.48  & 37.39  & 84.66  & 24.82  & 35.63  \\

		\midrule
		\multirow{8}{*}{\rotatebox[origin=c]{90}{Tiny ImageNet}} 
		
	& MIFGSM & 91.16  & 34.58  & 55.99  & 97.09  & 34.42  & 56.13  & 96.65  & 34.67  & 56.24  & 99.64  & 34.56  & 56.20  \\
      & CW    & 100.00  & 34.94  & 56.24  & 100.00  & 34.94  & 56.18  & 99.98  & 35.01  & 56.28  & 100.00  & 35.04  & 56.23  \\
      & TIMIFGSM & 72.83  & 35.01  & 56.26  & 73.96  & 35.08  & 56.30  & 85.73  & 34.94  & 56.34  & 92.09  & 34.92  & 56.28  \\
      & Wasserstein & 73.75  & 32.37  & 55.65  & 77.02  & 33.06  & 55.81  & 70.47  & 32.30  & 55.62  & 62.83  & 33.59  & 55.88  \\
      & cAdv  & 34.18  & 34.61  & 56.51  & 50.65  & 34.60  & 56.36  & 41.94  & 34.62  & 56.58  & 45.30  & 34.65  & 56.53  \\
          
      \cline{2-14}
      & \textbf{RI-AVA} & 21.56  & 34.06  & 55.53  & 25.54  & 34.22  & 55.76  & 22.18  & 33.97  & 55.60  & 33.77  & 33.99  & 55.87  \\
      & \textbf{RA-AVA} & 69.44  & 29.33  & 51.98  & 90.23  & 28.96  & 51.44  & 76.98  & 28.95  & 52.15  & 96.92  & 28.87  & 52.26  \\
		
		\bottomrule
		
	\end{tabular}
	}
	}
}
\caption{Comparison results on 3 datasets with 5 attack baselines and our methods. It contains the success rates (\%) of whitebox adversarial attack on three normally trained models: ResNet50, EfficientNet-b0, DenseNet121 and MobileNet-v2. The 1st column displays the whitebox attack results. The last two columns show the BRISQUE and NIQE score.}
\label{Tab-whitebox}
\end{table*}

\begin{figure*}[t]
	\centering
	\includegraphics[width=\textwidth]{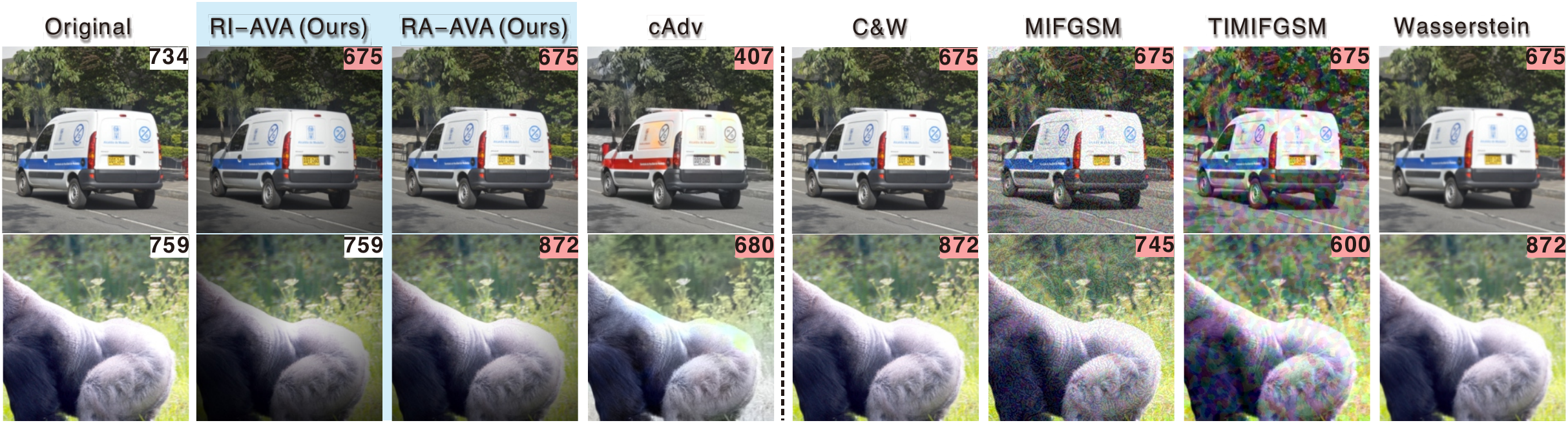}
	\caption{ Adversarial examples generated with different methods. The top right corner shows the predictive label index and the `red' numbers presents the attack misleads the DNN successfully.  RI-AVA, RA-AVA, and cAdv are non-noise-based adversarial attack.
	}
	\label{fig:visbaseline}
\end{figure*}

Here, we conduct comprehensive experiments on three popular datasets to evaluate the effectiveness of our method. We compare our method with some popular baselines including adversarial noise attack methods and other methods. Finally, we conduct experiments to showcase that our method can effectively defend against vignetting corrections.


{\bf Datasets.}
We carry out our experiments on three popular datasets, \ie, DEV \cite{devDB}, CIFAR10 \cite{krizhevsky2009learning}, and Tiny ImageNet \cite{tinyImageNet}.

{\bf Models.}
In order to show the effect of our attack method on different neural network models, we choose four popular models to attack, \ie, ResNet50 \cite{he2016deep}, EfficientNet-B0 \cite{tan2019efficientnet}, DenseNet121 \cite{huang2017densely}, and MobileNet-V2 \cite{sandler2018mobilenetv2}. We train these models on the CIFAR10 and Tiny ImageNet dataset. For DEV dataset, we use the pretrained models.

{\bf Metrics.}
We choose attack success rate and image quality to evaluate the effectiveness of our method. The image quality measurement metrics are BRISQUE \cite{mittal2012no} and NIQE \cite{mittal2012making}. BRISQUE and NIQE are two non-reference image quality assessment methods. A high score for BRISQUE or NIQE indicates poor image quality.

{\bf Baseline methods.}
We compare our method with five SOTA attack baselines, \ie, momentum iterative fast gradient sign method (MIFGSM) \cite{dong2018boosting}, Carlini \& Wagner L2 method (C\&W$_\text{L2}$) \cite{carlini2017towards}, translation-invariant momentum iterative fast gradient sign method (TIMIFGSM) \cite{dong2019evading}, Wasserstein attack via projected sinkhorn iterates (Wasserstein) \cite{wong2019wasserstein} and colorization attack (cAdv) \cite{bhattad2019unrestricted}.

\textbf{Analysis of physical parameters.}
Focal length and geometric factor are the two important parameters for the vignetting effect.
We evaluate the influence of the two physical parameters by setting different norm ball constraint for $f^{-1}$, $\alpha$ via Eq~\eqref{eq:riava_adv_obj}.
According to the result in Fig.~\ref{fig:f_alf_trend}(a) and (b), we observe that: \ding{182} Given different ball bound to $f$ and $\alpha$, the success rate of attack will be different. \ding{183} With the ball bound of $f$ growing, the success rate decreases. But as ball bound of $\alpha$ increases, the success rate increases. \ding{184} From the visualization results in Fig.~\ref{fig:f-alf}, with the value of $f$ decreasing and the value of $\alpha$ increasing, the vignetting effect becomes more obvious. Therefore, 
we can conclude that a stronger vignetting effect can increase the attack success rate.

\textbf{Analysis of different objective functions.}
We evaluate the effect of different energy terms by setting different coefficient values of energy terms, \ie, $\lambda_{g}$, $\lambda_{f}$ and $\lambda_{\alpha}$.
From the result in Fig.~\ref{fig:f_alf_trend}(c)(d), we can see that: \ding{182} 
With the increasing of $\lambda_{g}$ and $\lambda_{f}$, the success rate of attack decreases accordingly. \ding{183} The restriction on the energy item will reduce the success rate, indicating that the energy item has certain constraints on the attack.
In addition, we find that the change of $\lambda_{\alpha}$ almost does not affect the attack success rate, which shows that this energy item has minimal constraints on the attack.

%
\begin{table*}[t]
\centering
\setlength{\tabcolsep}{1.9pt} 
\tiny{
	\begin{tabular}{l|l|ccccc|ccccc|ccccc|ccccc}
		
		\toprule
		
		\rowcolor{tabgray}\cellcolor {white}&\multicolumn{1}{c|}{\tabincell{c}{Crafted from}} & \multicolumn{5}{c|}{ResNet50} & \multicolumn{5}{c|}{EfficientNet}& \multicolumn{5}{c|}{DenseNet} & \multicolumn{5}{c}{MobileNet} \\
		\cmidrule(r){2-2} \cmidrule(r){3-7} \cmidrule(r){8-12} \cmidrule(r){13-17}\cmidrule(r){18-22}
		\rowcolor{tabgray}
		\cellcolor {white}\multirow{-2}{*}{} & \tabincell{c}{Attacked model\\ \& Metrics} 
		   & EffNet & DenNet & MobNet & BRISQUE & NIQE
		
		   & ResNet & DenNet & MobNet & BRISQUE & NIQE
		
		   & ResNet & EffNet & MobNet & BRISQUE & NIQE
		   
		   & ResNet & EffNet & DenNet & BRISQUE & NIQE\\
		
		\midrule
		\multirow{8}{*}{\rotatebox[origin=c]{90}{DEV}} 
		& MIFGSM & \cellcolor{top3}13.29  & \cellcolor{top3}16.50  & \cellcolor{top3}12.81  & 20.93  & 39.42  & \cellcolor{top3}13.67  &\cellcolor{top3} 15.95  & \cellcolor{top2}31.26  & 19.12  & 39.38  & \cellcolor{top3}11.30  & \cellcolor{top3}9.75  & \cellcolor{top3}11.40  & 21.17  & 40.19  & \cellcolor{top2}7.97  & \cellcolor{top2}21.15  & \cellcolor{top3}10.08  & 22.79  & 42.19  \\
      & CW    & 0.55  & 0.78  & 0.12  & 17.49  & 48.36  & 0.32  & 1.11  & 2.00  & 17.81  & 48.36  & 0.11  & 0.33  & 0.12  & 17.48  & 48.53  & 0.00  & 0.33  & 0.22  & 17.33  & 48.51  \\
      & TIMIFGSM & 5.54  & 9.08  & 8.81  & 18.34  & 45.86  & 7.64  & 10.74  & 15.75  & 18.59  & 46.15  & 5.71  & 6.53  & 7.40  & 18.48  & 45.60  & 3.12  & 9.08  & 7.42  & 18.55  & 45.99  \\
      & Wasserstein & 0.78  & 1.99  & 1.53  & 20.91  & 51.63  & 0.65  & 1.11  & 1.65  & 20.44  & 51.30  & 0.86  & 0.66  & 1.06  & 20.67  & 51.79  & 0.32  & 0.33  & 0.44  & 20.02  & 51.58  \\
      & cAdv  & \cellcolor{top1}28.35  & \cellcolor{top1}34.33  & \cellcolor{top1}32.31  & 18.44  & 51.48  & \cellcolor{top1}31.93  & \cellcolor{top1}36.61  & \cellcolor{top1}45.77  & 18.45  & 51.38  & \cellcolor{top1}22.39  & \cellcolor{top1}23.37  & \cellcolor{top1}27.03  & 18.43  & 51.46  & \cellcolor{top1}25.83  & \cellcolor{top1}29.24  & \cellcolor{top1}30.79  & 18.44  & 51.36  \\
      
      \cline{2-22}
      & \textbf{RI-AVA} & 3.32  & 4.10  & 4.11  & 19.78  & 48.33  & 3.88  & 5.65  & 5.76  & 20.06  & 48.24  & 4.41  & 4.54  & 5.17  & 20.10  & 48.17  & 5.17  & 7.31  & 6.98  & 20.32  & 48.12  \\
      & \textbf{RA-AVA} & \cellcolor{top2}20.27  & \cellcolor{top2}21.59  &\cellcolor{top2} 23.97  & 21.33  & 46.92  & \cellcolor{top2}17.65  & \cellcolor{top2}20.93  & \cellcolor{top3}28.91  & 22.81  & 47.02  & \cellcolor{top2}20.02  & \cellcolor{top2}22.04  & \cellcolor{top2}23.38  & 20.89  & 46.54  & \cellcolor{top2}16.36  & \cellcolor{top2}22.48  & \cellcolor{top2}18.16  & 21.20  & 46.66  \\
		\midrule
		\multirow{8}{*}{\rotatebox[origin=c]{90}{CIFAR10}}  & MIFGSM & \cellcolor{top2}14.08  & \cellcolor{top2}13.09  & \cellcolor{top3}12.46  & 41.86  & 42.04  & \cellcolor{top3}10.26  & \cellcolor{top3}9.02  & \cellcolor{top2}17.46  & 41.90  & 42.53  & \cellcolor{top2}13.49  & \cellcolor{top2}13.40  & \cellcolor{top3}11.74  & 41.40  & 41.97  & \cellcolor{top3}4.46  & \cellcolor{top3}9.24  & \cellcolor{top3}4.23  & 41.56  & 42.14  \\
      & CW    & 5.01  & 2.75  & 5.76  & 41.43  & 41.36  & 1.04  & 1.11  & 3.96  & 41.66  & 41.06  & 2.51  & 4.78  & 5.19  & 41.34  & 41.42  & 0.54  & 1.07  & 0.43  & 41.46  & 41.01  \\
      & TIMIFGSM & 5.47  & 5.63  & 5.79  & 41.66  & 40.64  & 3.92  & 3.82  & 7.13  & 41.54  & 40.59  & 4.61  & 4.61  & 4.91  & 41.48  & 40.62  & 2.24  & 4.05  & 2.21  & 41.44  & 40.67  \\
      & Wasserstein & 8.25  & 2.65  & 5.82  & 45.45  & 44.47  & 0.86  & 0.71  & 2.55  & 44.81  & 44.09  & 2.34  & 8.10  & 5.03  & 43.26  & 42.05  & 0.52  & 1.25  & 0.35  & 43.45  & 42.87  \\
      & cAdv  & \cellcolor{top3}10.92  & \cellcolor{top3}9.16  & \cellcolor{top2}13.47  & 41.42  & 40.78  & \cellcolor{top2}11.54  & \cellcolor{top2}10.98  & \cellcolor{top3}17.34  & 41.55  & 40.78  & \cellcolor{top3}10.35  & \cellcolor{top3}10.87  & \cellcolor{top2}13.27  & 41.32  & 40.84  & \cellcolor{top2}9.28  &  \cellcolor{top2}10.91  & \cellcolor{top2}8.27  & 41.40  & 40.84  \\
      \cline{2-22}
      & \textbf{RI-AVA} & 2.61  & 2.62  & 2.18  & 40.54  & 40.28  & 2.94  & 2.93  & 2.92  & 39.90  & 40.34  & 2.96  & 2.56  & 2.26  & 40.57  & 40.28  & 2.73  & 3.28  & 2.87 & 39.73  & 40.36  \\
      & \textbf{RA-AVA}  & \cellcolor{top1}18.21  & \cellcolor{top1}19.18  & \cellcolor{top1}15.33  & 33.56  & 38.05  & \cellcolor{top1}25.42  & \cellcolor{top1}26.55  & \cellcolor{top1}30.43  & 28.29  & 35.39  & \cellcolor{top1}21.67  &\cellcolor{top1} 21.33  &\cellcolor{top1} 17.96  & 31.48  & 37.39  &\cellcolor{top1} 19.26  &\cellcolor{top1} 26.66  & \cellcolor{top1}19.31  & 24.82  & 35.63  \\
		
		\midrule
		\multirow{8}{*}{\rotatebox[origin=c]{90}{Tiny ImageNet}}   
    & MIFGSM & \cellcolor{top3}18.71  & \cellcolor{top3}22.81  & \cellcolor{top3}17.19  & 34.58  & 55.99  & \cellcolor{top3}13.37  & \cellcolor{top3}15.48  & \cellcolor{top3}23.30  & 34.42  & 56.13  & \cellcolor{top3}17.74  & \cellcolor{top3}18.43  & \cellcolor{top3}15.73  & 34.67  & 56.24  & \cellcolor{top3}5.10  & \cellcolor{top3}9.61  & \cellcolor{top3}5.40  & 34.56  & 56.20  \\
      & CW    & 5.24  & 4.90  & 5.87  & 34.94  & 56.24  & 2.07  & 2.07  & 5.83  & 34.94  & 56.18  & 2.48  & 3.30  & 3.83  & 35.01  & 56.28  & 0.45  & 0.90  & 0.38  & 35.04  & 56.23  \\
      & TIMIFGSM & 10.44  & 15.06  & 11.76  & 35.01  & 56.26  & 7.66  & 9.89  & 14.88  & 35.08  & 56.30  & 10.34  & 10.22  & 10.99  & 34.94  & 56.34  & 4.33  & 6.45  & 4.97  & 34.92  & 56.28  \\
      & Wasserstein & 5.81  & 4.97  & 7.75  & 32.37  & 55.65  & 2.13  & 2.37  & 5.99  & 33.06  & 55.81  & 2.94  & 3.84  & 5.22  & 32.30  & 55.62  & 0.70  & 0.92  & 0.63  & 33.59  & 55.88  \\
      & cAdv  & \cellcolor{top2}28.24  & \cellcolor{top2}29.53  & \cellcolor{top1}34.57  & 34.61  & 56.51  & \cellcolor{top1}31.40  & \cellcolor{top1}32.93  & \cellcolor{top1}42.37  & 34.60  & 56.36  & \cellcolor{top2}27.04  & \cellcolor{top2}28.19  & \cellcolor{top1}34.51  & 34.62  & 56.58  & \cellcolor{top1}25.13  & \cellcolor{top1}28.12  & \cellcolor{top1}26.31  & 34.65  & 56.53  \\
      \cline{2-22}
      & \textbf{RI-AVA} & 7.02  & 8.10  & 6.98  & 34.06  & 55.53  & 6.63  & 6.78  & 8.87  & 34.22  & 55.76  & 7.87  & 7.16  & 6.44  & 33.97  & 55.60  & 6.28  & 8.44  & 5.85  & 33.99  & 55.87  \\
      
      & \textbf{RA-AVA} &\cellcolor{top1} 30.45  & \cellcolor{top1}32.42  &\cellcolor{top2} 29.72  & 29.33  & 51.98  & \cellcolor{top2}27.41  & \cellcolor{top2}28.20  & \cellcolor{top2}37.61  & 28.96  & 51.44  & \cellcolor{top1}29.49  & \cellcolor{top1}29.98  & \cellcolor{top2}29.53  & 28.95  & 52.15  & \cellcolor{top2}19.56  & \cellcolor{top2}25.68  & \cellcolor{top2}19.82  & 28.87  & 52.26  \\
    
		\bottomrule
		
	\end{tabular}
	}
\caption{Adversarial comparison results on three datasets with five attack baselines and our methods. It contains the success rates (\%) of transfer adversarial attack on three normally trained models: ResNet50 (ResNet), EfficientNet-b0 (EffNet), DenseNet121 (DenNet), and MobileNet-v2 (MobNet). The first three columns display the transfer attack results, where we use red, yellow, and blue to mark the first, second, and third highest success rate. And the last two columns show the BRISQUE score and NIQE score. 
}
\label{Tab-tranfer}
\end{table*}
%


\textbf{Comparison with baseline attacks.}
We evaluate the effectiveness of our attack with the baseline.
Table~\ref{Tab-whitebox} shows the quantitative results about the attack success rates and the image quality.
We observe that for every dataset, RA-AVA reaches higher success rate than RI-AVA, indicating that tunable vignetting regions can greatly improve the success rate of attack.
Compared with the noise-based adversarial attack, we found that our attack achieved lower success rate than MIFGSM, CW and TIMIFGSM attack. This is in line with our expectation since the image vignetting has more constraints on image perturbation than adding arbitrary noises. This is also why our method (RA-AVA) could have better image quality than the noise-based attack. We also noticed that on some models and datasets (\eg, DEV), RA-AVA could still achieve competitive results in terms of attack success rate while the image quality is better. Compared with the non-noise based attack, our method is better than cAdv significantly.
Furthermore, we also find that our method can achieve much better transferability, which will be introduced later.

%

In Fig.~\ref{fig:visbaseline}, we have showcased some examples generated by baselines and our attack methods. The first column shows the original images while the following columns list the corresponding adversarial examples. It is clear that our method could generate high-quality adversarial examples that are smooth and realistic. However, we could find obvious noises in the examples generated by the adversarial noise attack methods, which are difficult to appear in the real world. For other non-noise attack methods, \eg, cAdv, they allow patterns that may appear in the real world but the change between the original and the generated image is very perceptible. Our method does not change the image too much while maintaining the realism in optical system for the vignetting effects.

\textbf{Comparison on transferability.}
We then evaluate the transferability of different attacks.
Table~\ref{Tab-tranfer} shows the quantitative transfer attack results of our methods and the baseline methods. In transfer attack, one attacks the target DNN with the adversarial examples generated from other models.
As we can see, in most cases, our method achieves much higher transfer success rate than others while the image quality is also higher. For example, the attack examples crafted from ResNet50 on DEV dataset achieves 37.21\%, 40.75\%, and 40.89\% transfer success rate on EfficientNet, DenseNet, and MobileNet, with the lowest values of BRISQUE and NIQE, \ie, 11 and 37.04. 

\begin{table}[htbp]
\centering
\setlength{\tabcolsep}{3pt}
\scriptsize
\begin{tabular}{l|rrrr}
    \toprule
    & \multicolumn{1}{l}{ResNet50} & \multicolumn{1}{l}{EfficientNet} & \multicolumn{1}{l}{DenseNet} & \multicolumn{1}{l}{MobileNet} \\
    \hline
    original & 66.06  & 57.83  & 65.33  & 49.40  \\
    RA-AVA & 19.72  & 4.47  & 14.60  & 0.71  \\
    zero-dce & 29.40  & 12.96  & 24.81  & 10.24  \\
    \bottomrule
\end{tabular}%
\caption{Accuracy of four models on Tiny Imagenet before attack, after RA-AVA attack and after Zero-DCE correction.}
\label{Tab-zerodce}%
\end{table}%
\begin{figure}[t]
	\centering
	\includegraphics[width=\columnwidth]{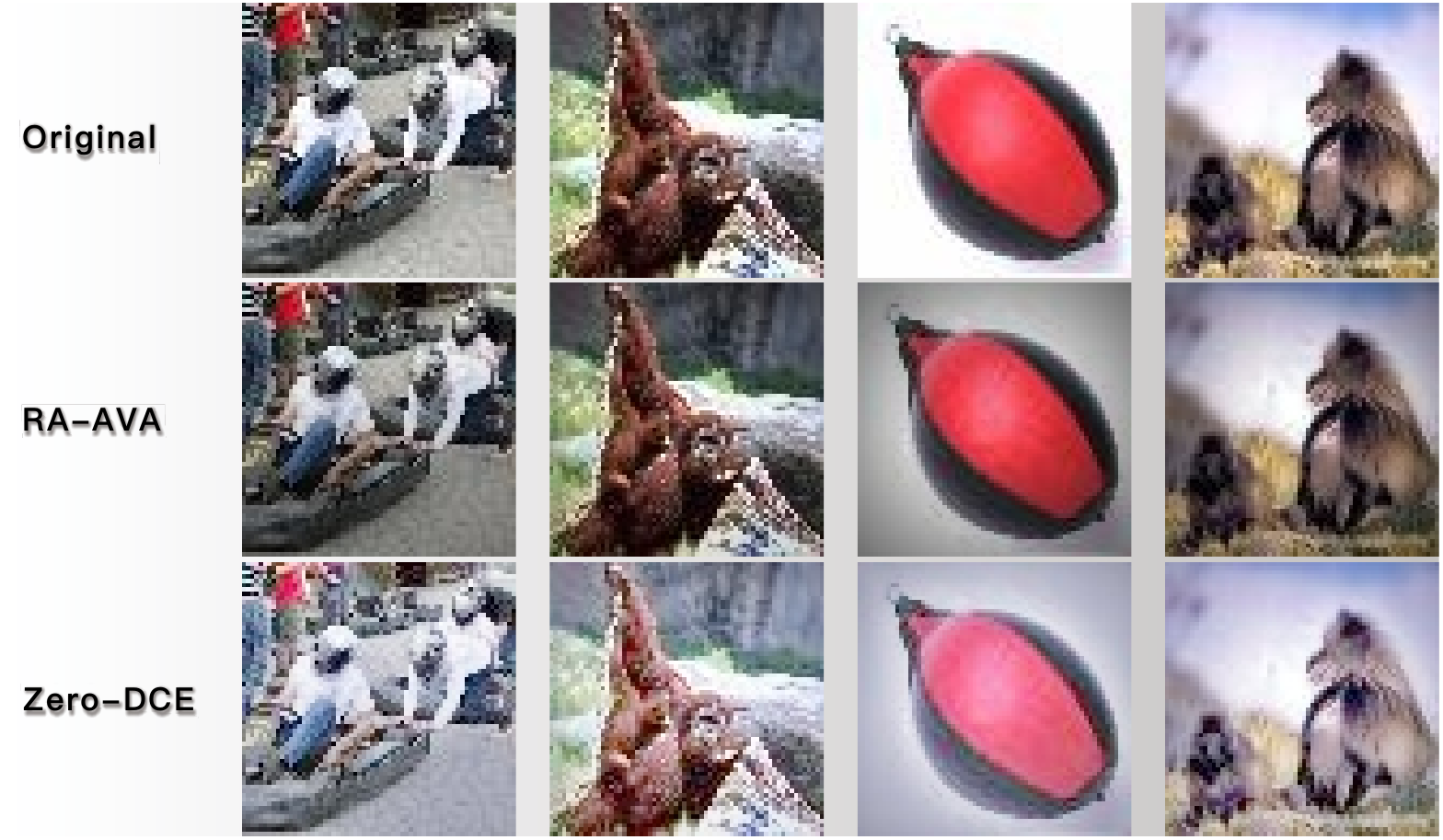}
	\caption{The visualization result of examples before attack, after RA-AVA attack and after Zero-DCE correction.}
	\label{fig:zerodce}
\end{figure}
\textbf{AVAs against vignetting corrections.}
Since our method is to use the vignetting effect as the attack method, we need to consider whether the method of light intensity and vignetting correction can neutralize our attack. For this reason, we use the Zero-DCE method \cite{guo2020zero} to adjust the light intensity. The visualization result is shown in Fig.~\ref{fig:zerodce}. It can be seen that the Zero-DCE method has performed a certain brightness correction on the attacked images. The quantitative results of the accuracy change are shown in 
Table~\ref{Tab-zerodce}. After Zero-DCE correction, the accuracy has a certain improvement, but it is still lower than the original. It shows that the Zero-DCE method could mitigate the attack on some images but it is still not effective (\eg, only about 10\% improvement), indicating that our attack method is robust against intensity and vignetting correction methods.

\section{Conclusion}

We have successfully embedded stealthy adversarial attack into the image vignetting effect through a novel adversarial attack method termed adversarial vignetting attack (AVA). By first mathematically and physically model the image vignetting effect, we have proposed the radial-isotropic adversarial vignetting attack (RI-AVA) and tuned the physical parameters such as the illumination factors and the focal length through the guidance of the target CNN models under attack. Next, by further allowing the effective regions of vignetting to be radial-anisotropic and shape-free, our proposed radial-anisotropic adversarial vignetting attack (RA-AVA) can reach much higher transferability across various CNN models. Moreover, level-set-based optimization is proposed to jointly solve the adversarial vignetting regions and physical parameters. 

The proposed AVA-enabled adversarial examples can fool the SOTA CNNs with high success rate while remaining imperceptible to human. Through extensive experiments on three popular datasets and via attacking four SOTA CNNs, we have demonstrated the effectiveness of the proposed method over strong baselines. We hope that our study can mark one small step towards a fuller understanding of adversarial robustness of DNNs. In a long run, it can be important to explore the interplay between the proposed adversarial vignetting attack and other downstream perception tasks that are usually mission critical such as robust tracking \cite{guo2020spark,cheng2021deepmix}, robust autonomous driving \cite{li2021fooling}, and robust DeepFake detection \cite{qi2020deeprhythm,arxiv21_dfsurvey}, \etc.


\textbf{Acknowledgement.}
This work has partially been sponsored by the National Science Foundation of China (No. 61872262). It was  supported in part by Singapore National Cyber-security R\&D Program No. NRF2018NCR-NCR005-0001, National Satellite of Excellence in Trustworthy Software System No. NRF2018NCR-NSOE003-0001, NRF Investigatorship No. NRFI06-2020-0022. We gratefully acknowledge the support of NVIDIA AI Tech Center (NVAITC) to our research.
{\small
\bibliographystyle{named}
\bibliography{ref}
}
\end{document}